\documentclass[3p,times,procedia]{elsarticle}
\usepackage{nupha_ecrc}

\volume{00}

\firstpage{1}

\journalname{Nuclear Physics A}

\runauth{ M. Gyulassy et al.}

\jid{nupha}

\jnltitlelogo{Nuclear Physics A}

\usepackage{amssymb}

\usepackage[figuresright]{rotating}
\usepackage{url}
\usepackage{hyperref}

\begin{document}

\begin{frontmatter}



\dochead{XXVIIIth International Conference on Ultrarelativistic Nucleus-Nucleus Collisions\\ (Quark Matter 2019)}

\title{Dijet Acoplanarity in CUJET3 as a Probe of the Nonperturbative Color Structure of QCD Perfect Fluids}


\author[LBL,CCNU]{M. Gyulassy}
\author[LBL]{P.M. Jacobs}
\author[IU]{J. Liao}
\author[McGill]{S. Shi}
\author[LBL,CCNU]{X.N. Wang}
\author[LBL]{F. Yuan}

\address[LBL]{Nuclear Science Division, Lawrence Berkeley National Laboratory, Berkeley, CA 94720, USA}

\address[CCNU]{Institute of Particle Physics, Central China Normal University, Wuhan, China}


\address[IU]{Physics Department and Center for Exploration of Energy and Matter,
Indiana University, \\ 2401 N Milo B. Sampson Lane, Bloomington, IN 47408, USA}

\address[McGill]{Dept. of Physics, McGill University,  
  3600 rue University, Montreal, QC H3A 2T8, Canada}

\address{}

\begin{abstract}
  Using the CUJET3=DGLV+VISHNU jet-medium interaction framework,
  we show that dijet azimuthal acoplanarity in high energy $A+A$ collisions
  is sensitive to possible non-perturbative enhancement of the jet transport coefficient, $\hat{q}(T,E)$, 
  in the QCD crossover temperature $T\sim 150-300$ MeV range.
  With jet-medium couplings constrained by global RHIC\& LHC $\chi^2$ fits to 
   nuclear modification data on $R_{AA}(p_T>20)\;{\rm GeV}$, 
   we compare predictions 
of the  medium induced
  dijet transverse momentum squared, 
  $Q_s^2\sim \langle \hat{q}L\rangle \sim   \Delta\phi^2 E^2$,
%
  in two models of the temperature, $T$,  and jet energy $E$  
  dependence of the jet medium transport coefficient, $\hat{q}(T,E)$.
  In one model, wQGP, the  chromo degrees of freedom (dof)
  are approximated by 
  a perturbative dielectric gas of quark and gluons dof.
  In the second model, sQGMP, we consider a nonperturbative
  partially confined semi-Quark-Gluon-Monopole-Plasma
  with emergent color magnetic dof constrained by lattice QCD data.
  Unlike the slow variation of the scaled jet transport coefficient, $\hat{q}_{wQGP}/T^3$,  the sQGMP model $\hat{q}_{sQGMP}/T^3$ features  a sharp maximum
  in the QCD confinement crossover $T$ range.
  We show that the dijet path averaged medium induced azimuthal
  acoplanarity, $\Delta\phi^2$, in sQGMP is robustly $\sim 2$ times
  larger than in perturbative wQGP. even though the radiative energy loss
  in both models is very similar as needed to fit the same $R_{AA}$ data.
  Future A+A dijet acoplanarity measurements {\em constrained} together
  with single jet $R_{AA}$ {\em and}  $v_n$ measurements
  therefore appears to be a promising strategy  
  to search for possible signatures  of critical 
  opalescence like phenomena in the QCD confinement temperature range.     
  \end{abstract}


\end{frontmatter}


\section{Introduction and Conclusions}
\label{intro}


Dijet relative azimuthal angle acoplanarity, $\Delta \phi^2=(\pi-\phi_1+\phi_2)^2= (Q_{vac}^2+Q_s^2)/E^2$,
as a probe of Quark Gluon Plasmas (QGP) produced in high energy nuclear collisions, has a long history, see e.g. \cite{Appel,Baier:1996sk,DEramo:2018eoy,Majumder:2008zg,Luo:2018pto, Gyulassy:2018qhr,Shi2020}. In both $p+p$ and $A+A$ collisions this observable
is always dominated by $Q_{vac}^2/E^2\sim \alpha_s^2\sim 0.1$ due
to (Sudakov) multiple gluon emission into the vacuum associated with
all hard QCD processes\cite{Ellis:1980my}.

The BDMS\cite{Baier:1996sk} medium dependent  
``saturation'' scale, $Q_s^2=\langle \hat{q}L\rangle =\langle \int dt \;\hat{q}(T(t))\rangle$, is a jet path averaged measure of jet straggling transverse to its initial direction $\hat{\mathbf n}(\phi_0)$. It is 
the average transverse momentum squared accumulated over a path length $L$.
For a jet with color,flavor $a$, the jet transport coefficient $\hat{q}_a$
in a  QCD
fluid of temperature $T=T({\mathbf x},t)$, 
 $\hat{q}_a(T)=\langle q^2/\lambda \rangle_a
=\int dq^2\; q^2\Gamma_a(q^2,T)$. It
depends on the composition densities ,
$\{\rho_b(T);b=q,g,m,\cdots\}$,
of effective color electric and magnetic
degrees of freedom (dof) as well as  on the 
microscopic differential scattering rates,
$\Gamma_a(q^2,T)\equiv
\sum_b \rho_b(T)  d\sigma_{ab}(T)/dq^2$.

In fact,  $Q_s^2(\hat{\mathbf n},E_{fin})=\langle
\int dt \;\hat{q}(T({\mathbf z}(t),t),E_{ini}) \rangle_{\{E_{ini},{\mathbf z}(0)\}}$,
is only one of a large set of jet path line integral functionals depending on jet paths, ${\mathbf z}(t)={\mathbf x}_0+ t \;\hat{\mathbf n}(\phi_0)$
that control the medium modification of jets in $A+A$.
In particular, the correct geometric averaging of $\langle Q_s^2(E_{fin})\rangle$
for given observed jet energy and direction
requires simultaneous calculation of radiative and elastic
energy loss functionals as well: 
$\Delta E_{rad}\sim \int dt\; t\;\hat{q}(t)\;{\cal F}_{rad}(t,E)$ and $\Delta E_{el}\sim \int dt\; \hat{q}(t)/T(t)$. This is because jet quenching strongly biases
 the spatial
distribution of jet initial production points, ${\mathbf x}_0$, to
 a sub region of the medium transverse geometry from which jets in a given direction $\hat{n}(\phi_0)$  emerge with given final energy $E_{fin}$. Note that 
 in the asymptotic $E\rightarrow \infty$ BDMS limit $\Delta E_{rad}^{BDMS}\sim
 \int dt\;t\hat{q}\approx \langle \hat{q}L^2\rangle/2$ because
 ${\cal F}_{BDMS}(t,\infty)\equiv1$.
 However, for non-asymptotic $E<100$ GeV jet energies of interest here,
 the  DGLV\cite{Gyulassy:2002yv} formalism predicts that
 $\Delta E_{rad}^{DGLV}=\int dt
 \int d^2{\mathbf q}\;\Gamma_a({\mathbf q},T(t))\;\left\{\int dxd^2{\mathbf k}\;
 A({\mathbf q},{\mathbf k},M^2(x,T) )\left((1-
 \cos[ t (({\mathbf k}-{\mathbf q})^2+M^2(x,T))/(2 x E)]\right)\right\}$.
 We found numerically that 
 only in the  high energy ($E>100$ GeV) limit can we approximate ${\cal F}_{rad}(t,E>100)\approx 1$.
 In the CUJET framework the DGLV energy loss integrals $\int dt d^2{\mathbf q}dx{\mathbf k}\cdots$  needed
 to compute $\Delta E_{rad}^{DGLV}\ne
 \int dt\;t\hat{q}_{DGLV}$ and hence $\Delta E_{rad}^{DGLV}$ does not simply scale
for moderate energy jets with $\hat{q}$ as does $Q_s^2[elas]$ by definition.

As emphasized in \cite{Appel} long ago, dijet acoplanarity , as a stand alone observable 
cannot uniquely discriminate between
 different models of the color dof $\rho_b(T)$ and
the microscopic $d\sigma_{ab}$ cross sections. 
This ambiguity is  further amplified
by the strong dependence of all jet path functionals 
on the  non static, inhomogeneous, anisotropic 
temperature field , $T({\mathbf z},t)$ produced in finite $A+A$ collisions.
 The unavoidable geometric bias caused by jet quenching implies that
the triple set of hard jet observables
$\{R_{AA},\;v_n,\;\Delta\phi\}$ must be strongly correlated. Hence, measuring the correlation between these three observables
should enhance the discriminating power of hard jet and dijet observables
to probe the color structure of QCD fluids,
as we emphasized in \cite{Gyulassy:2018qhr,MGQM19}.

Current interest in $A+A$ dijet acoplanarity observables is motivated by the
first preliminary data from RHIC\cite{phenixstar,STAR17} and LHC\cite{ALICE15}
that suggest \cite{Mueller:2016gko,Chen:2016vem}
that future higher statics measurements of the acoplanarity distribution
in the ``sweet spot''
$20<E_{fin}<80$ GeV jet energy range 
will be able to resolve   
medium induced corrections, $\Delta \phi_{med}^2=Q_s^2/E^2 $ 
from the dominant Sudakov source of dijet
acoplanarity\cite{Ellis:1980my} that can be directly measured in $p+p$.

Another important  motivation for our 
focus on dijet acoplanarity here is that there  exist currently
several independent frameworks\cite{Jaki,Shen:2014vra,Niemi:2015qia} with rather different
combinations of quenching dynamics and viscous hydrodynamics modeling
that have tested equally well at the $\chi^2/dof<2$ level against
currently available soft {\em and} hard  $R_{AA}$ and $v_n$ data in $A+A$ at RHIC and LHC.
This work is thus also motivated 
by the question: ``Can dijet acoplanarity 
help experimentally to break
the current degeneracy between soft+hard modeling of $A+A$?''

The CUJET3 framework used here utilized the
temperature and flow velocity fields predicted by VISHNU2+1 \cite{Shen:2014vra} code with Glauber Initial Conditions.
The  DGLV jet quenching
theory\cite{Gyulassy:2002yv} is then applied to evaluate both 
$\Delta E$ and $Q_s^2$ jet path functionals in 
the VISHNU2+1 viscous hydrodynamic fluid fields.
See Refs.\cite{Shi:2018vys,CIBJET,Shi:2017onf,Xu:2015bbz,Xu:2014tda,Xu:2014ica}
for details.  Our global $\chi^2/dof <2$  fit\cite{Shi:2018vys} to
available soft+hard data constrained the two free parameters of CUJET3:
the maximum of the running coupling $\alpha_c=0.9\pm 0.1$ and the ratio of magnetic to electric screening scales, $c_m=\mu_M(T)/(g(T)\mu_D(T)) \approx 0.25\pm 0.03$. We use the same values of the two parameters to compute $Q_s^2(E)$ here. 

The CUJET3 jet path functionals are evaluated 
in two models, wQGP and sQGMP, of the color composition  of the QCD fluid .
The wQGP composition model assumes the color structure of the QCD fluid
can be approximated by perturbative 
two component color di-electric model with one loop dynamically screened quark and gluon dof. However, for consistency with lattice QCD equation of state,
the Stefan-Boltzmann partial pressures,
$P_{b}^{SB}(T)=T\rho^{SB}_b(T)$, 
are scaled down by the ratio of the nonperturbative lattice QCD pressure, $P_{lat}(T)$, to the ideal gas pressure $T(\rho^{SB}_q+\rho^{SB}_g)$.

The sQGMP composition model includes emergent color magnetic
monopole (cmm) degrees of freedom, as proposed in \cite{Liao:2006ry,Liao:2008jg,Liao:2008dk} to solve the $R_{AA}\times v_2$ puzzle.
In CUJET3, the sQGMP model
further generalizes wQGP by not only adding the monopole dof
but also
by further reducing the q and g  dof partial pressures
by powers of the nonperturbative lattice Polyakov loop, $L(T)$, and/or
the light quark susceptibility, $\chi^u_s(T)$, data as proposed by \cite{Hidaka}.
 See \cite{Shi:2018vys,CIBJET,Shi:2017onf,Xu:2015bbz,Xu:2014tda,Xu:2014ica} for further details.

In Fig.1 (Left panel) we plot the 
the global $R_{AA}$ $\chi^2$ data constrained quark
jet transport fields, $\hat{q}_{sQGMP}(T,E)$ and $\hat{q}_{wQGP}(T,E)$.
Note that $\hat{q}_{sQGMP}(T,E)$ is strongly enhanced relative to
$\hat{q}_{wQGP}(T,E)$
in the QCD crossover temperature range $160<T<320$ MeV. This is  due to enhanced
jet-monopole interactions with
$d\sigma_{qm} \propto 
\alpha_E\alpha_M =1 \gg d\sigma_{qg}\propto \alpha_E^2 $. 
The question addressed here is whether $R_{AA}$ constrained
acoplanarity 
could serve to search for such 
``critical opalescence'' like  phenomena near the confinement temperature range.
Our answer is positive, as we show below.

In the Middle panel of Fig.1, the spacetime isochrone evolution of the
VISHNU temperature field in central 0-10\% $Pb+Pb$ 5 ATeV is shown. In the Right
panel of Fig.1 the
isochronous evolution of $\hat{q}_{wQGP}$ and
$\hat{q}_{sQGMP}$ are compared as a function of the reaction plane
$x$ coordinate with at  $y=0$.
The emergent monopole degrees in the crossover temperature
range near the freeze-out surface $T=160$ MeV  and at late times
are seen to enhance $\hat{q}_{sQGMP}$ by a factor $\sim 4$.
The enhancement of $\hat{q}$ near the crossover
surface regions  plays the decisive role , as proposed 
in \cite{Liao:2006ry,Liao:2008jg,Liao:2008dk}, in enhancing the CUJET3
predicted elliptic azimuthal asymmetry, $v_2$, in agreement with data. 
.

Our main new result shown in Fig.2 is that with charged hadron $R_{AA}$ constrained $\hat{q}(x,t)$ transport fields, the medium induced single jet acopanarity broadenning $\Delta\phi^2$ is robustly
a factor of $\sim 2$ larger in a QGP fluid with magnetic monopole degrees of freedom
than in a purely  di-electric (pQCD/HTL type) ``wQGP'' fluid.
.

\begin{figure}[!hbt]
  \begin{center} \vspace{-0.1in}
 \includegraphics[width=0.25\textwidth]{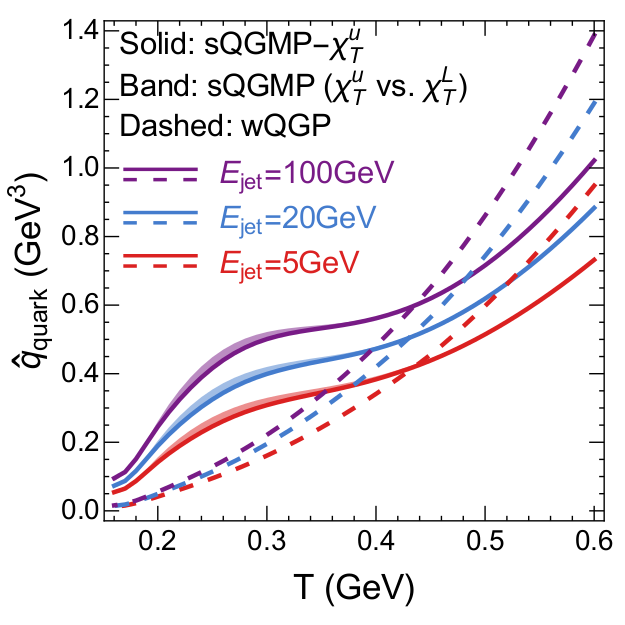}
\hspace{.2in} \includegraphics[width=0.25\textwidth]{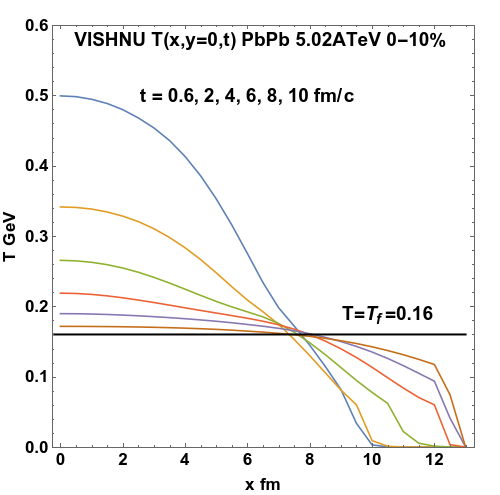}
\hspace{.2in} \includegraphics[width=0.25\textwidth]{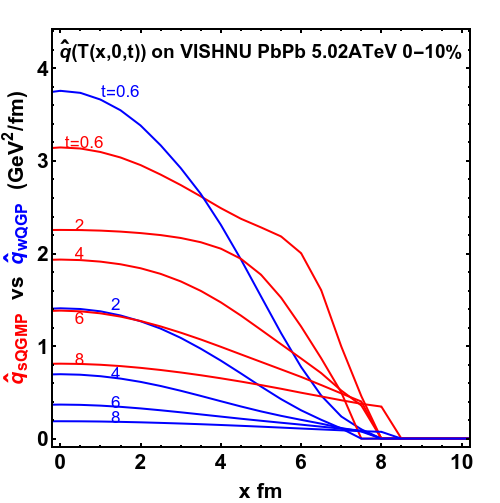}
 \vspace{-0.15in}
 \caption{(color online) (Left) The CUJET3.1 $R_{AA}$ constrained \protect \cite{Shi:2018vys,CIBJET,Shi:2017onf,Xu:2015bbz}
   jet transport field, $\hat{q}_F(T,E)$ for quark jets with
   $E_{ini}=5,20,100$ GeV  are compare  to wQGP and sQGMP
   models of the chromo electric and magnetic dof
  in the QCD fluid.
  Dashed curves for wQGP assume only color di-electric dof
  while solid curves for sQGMP assume that the color electric quark and gluon dof
  are suppressed by lattice Polyakov loop 
  and quark susceptibility factors, $\chi_T^u$,
  due to only partial confinement in $160<T<320$ MeV QCD transition range.
  In sQGMP the remaining dof are assumed to be color magnetic monopole quasi-parton dof.
  (Center) The isochronous evolution of 
  temperature field, $T(x,0,t)$,from  VISHNU2+1 viscous hydrodynamics\protect \cite{Shen:2014vra} for 0-10\% $Pb+Pb$ 5.02ATeV is shown.
  (Right) The isochronous evolution of the jet transport coefficients, $\hat{q}_{wQGP}$ (Blue) and $\hat{q}_{sQGMP}$ (Red), for $E=20\;{\rm GeV}$ are compared
  at given $x, y=0$ at times $t=0.6, \cdots,10$ fm/c . 
  Note that $\hat{q}_{sQGMP}$ is strongly  enhanced compared to  $\hat{q}_{wQGP}$
  in the surface regions
  and in interior at  late times. }

\label{fg1}

\end{center}
\vspace{0.in}
\end{figure}

\begin{figure}[!hbt]
 \begin{center} \vspace{-0.1in}
\includegraphics[width=0.25\textwidth]{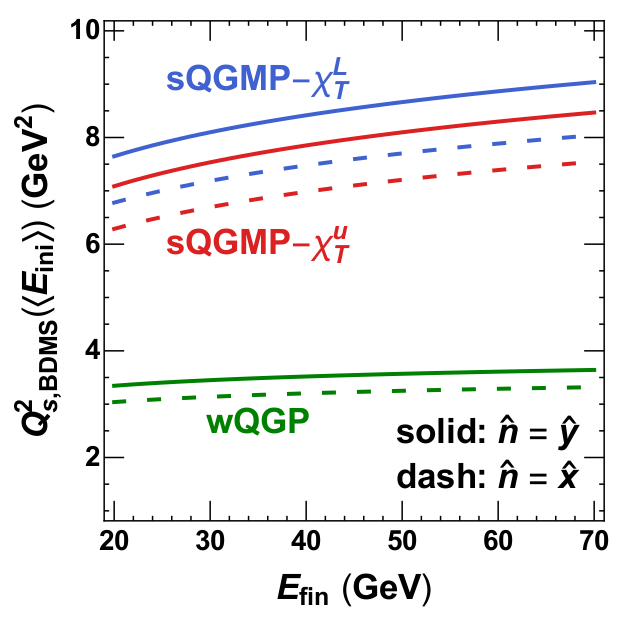}
\hspace{.2in}\includegraphics[width=0.25\textwidth]{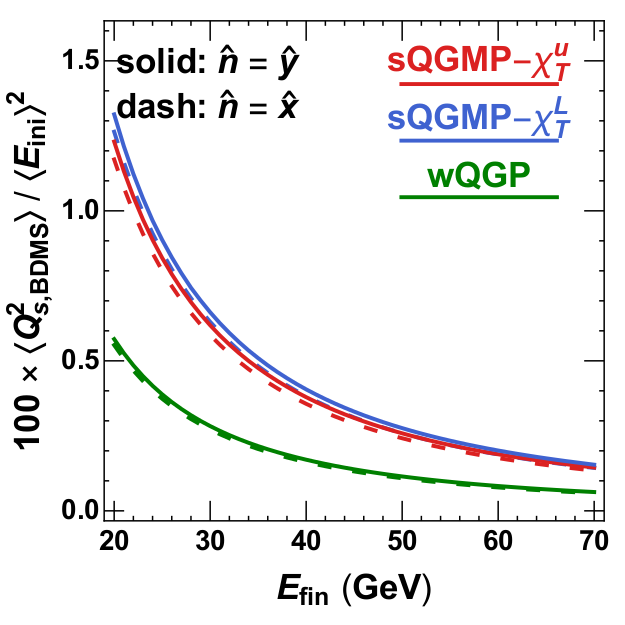}
\vspace{0.in}
\caption{(Left) Comparison of $R_{AA}^{ch}$ constrained
  CUJET3.1 predictions for single parton jet level  $Q_s^2$ in 20-30\% centrality Pb+Pb 5.02 ATeV.
  The final quenched energy , $E_{fin}$, dependence of the average  BDMS transverse momentum squared scale is compared for the three
  models of the color structure of QCD fluids as in Fig.1a but
  using the evolving VISHNU fluid $T(x,t)$ filed 20-30\% centrality class.
  Green curves show predictions in wQGP fluids, blue curves show sQGMP with semi quark and and semi gluon degrees of freedom suppressed by powers, $L(T)^{1}$ and $L(T)^2$ resp., of lattice Polyakov loop data. The red curves show results
  assuming an sQGMP fluid with semi quarks suppressed by lattice
  light quark susceptibility data on $\chi^u_2(T)$,
  while semi-gluons are suppressed by $L(T)^2$ (see \protect\cite{Shi:2018vys} for details). 
  (Right) CUJET3.1 predictions for 
  medium induced azimuthal angle broadening width squared $\Delta\phi^2= Q_{s}^2/\langle E_\mathrm{ini}\rangle^2$ at the single parton level
  averaged over both q and g jets. The abscissa is scaled up by a factor 100
for clarity. 
}

\label{fg2}
\end{center}
\vspace{-0.3in}
\end{figure}

Our previous study\cite{Gyulassy:2018qhr} of dijet acoplanarity 
concentrated on the tails of the $dN/\Delta \phi$
distributions in the $2.4<
\Delta \phi<3 $ range and showed that future experiments must
reach sub-percent levels of precision
to discriminate between medium dependent 
BDMS Gaussian and  DGLV, power law like Rutherford tails
convoluted on top of the dominant Sudakov vacuum radiation tails.
 The present study\cite{MGQM19}, summarized in Fig.2, 
utilized the CUJET3=DGLV+VISHNU 
framework \cite{Shi:2018vys,CIBJET,Shi:2017onf,Xu:2015bbz}
to compute, at  leading partonic level, the elastic 
$Q_s^2[E_{fin}]=\langle \int dt \hat{q}(T(t),E_{fin} +\Delta E({\mathbf x}_0 , \phi_0)\rangle$,
  taking into account the unavoidable geometric ''sunny side up'' bias
  due to jet energy loss via  $\Delta E({\mathbf x}_0 , \phi_0)
  = \Delta E_{rad}+\Delta E_{elas}$ , that we 
  constrained by global fits to data on nuclear modification
  os high $p_T$ hadron fragments,  $R_{AA}^{ch}(p_T)$.
We compared different models of the temperature dependence of the
color dof composition of the QCD fluid constrained not only by 
$R_{AA}^{ch}$ data but also by numerical lattice QCD equation of state data.


Our main new result, shown in Fig.2a, is that elastic scattering
$Q_s^2$ is predicted to be robustly $\sim 2$ times larger
in sQGMP than in wQGP models of the color structure\cite{Shi2020}.
Future work must next resolve 
the current debate on the sign and magnitude of
radiative corrections, $\Delta Q_s^2[rad]$
to elastic $Q_s^2[elas]$ \cite{Blaizot:2019muz,Zakharov:2019fov,CUJETrad}.
Our preliminary estimates, to be reported elsewhere \cite{CUJETrad}, agree with
Ref. \cite{Zakharov:2019fov} that $\Delta Q_s^2[rad]$
reduces moderately the magnitude of elastic scattering induced acoplanarity.

\vspace{-0. in}
{\small {\bf Acknowledgments.} 
  SS is supported by the Natural Sci. and Engin.Research Council of Canada,
  JL by 
  NSF PHY-1913729,  XNW and FY by  DOE DE-AC02-05CH11231, PJ
  by DOE Contract DE-AC02-05CH1123, and
 MG and XNW by 
 NSFC grants 11935007, 11221504 and 11890714.
 } 
\vspace{-0.3in}

\bibliographystyle{elsarticle-num}
\bibliography{<your-bib-database>}


\end{document}